\documentclass[%
 reprint,
superscriptaddress,
nofootinbib,
 amsmath,amssymb,
 aps,
]{revtex4-1}
\usepackage[colorlinks,linktocpage,linkcolor=cyan,citecolor=cyan]{hyperref}
\usepackage{graphicx}
\usepackage{dcolumn}
\usepackage{bm}
\usepackage{comment}
\usepackage{amsfonts,amssymb,mathtools,graphicx,color,bm}
\definecolor{ultramarine}{rgb}{0.07, 0.04, 0.56}
\definecolor{cadmiumgreen}{rgb}{0.0, 0.42, 0.24}
\definecolor{indigo(dye)}{rgb}{0.0, 0.25, 0.42}

\newcommand{\de}{\, \mathrm{d}}

\begin{document}
\begin{flushright} {\footnotesize YITP-24-61} \end{flushright}

\affiliation{School of Physics Science and Engineering, Tongji University, Shanghai 200092, China}
\affiliation{Institute for Advanced Study of Tongji University, Shanghai 200092, China}

\author{Misao Sasaki}
\email{misao.sasaki@ipmu.jp}
\affiliation{Kavli Institute for the Physics and Mathematics of the Universe (WPI), UTIAS, The University of Tokyo, Chiba 277-8583, Japan}
\affiliation{Asia Pacific Center for Theoretical Physics, Pohang 37673, Korea}
\affiliation{Center for Gravitational Physics and Quantum Information, Yukawa Institute for Theoretical Physics, Kyoto University, Kyoto 606-8502, Japan}
\affiliation{Leung Center for Cosmology and Particle Astrophysics, National Taiwan University, Taipei 10617, Taiwan}

\author{Vicharit Yingcharoenrat}
\email{Corresponding author: vicharit.y@chula.ac.th}
\affiliation{High Energy Physics Research Unit, Department of Physics, Faculty of Science, Chulalongkorn University, Pathumwan, Bangkok 10330, Thailand}
\affiliation{Kavli Institute for the Physics and Mathematics of the Universe (WPI), UTIAS, The University of Tokyo, Chiba 277-8583, Japan}

\author{Ying-li Zhang}
\email{yingli@tongji.edu.cn}
\affiliation{School of Physics Science and Engineering, Tongji University, Shanghai 200092, China}
\affiliation{Institute for Advanced Study of Tongji University, Shanghai 200092, China}
\affiliation{Institute of Theoretical Physics, Chinese Academy of Sciences, Beijing 100190, China}
\affiliation{Kavli Institute for the Physics and Mathematics of the Universe (WPI), UTIAS, The University of Tokyo, Chiba 277-8583, Japan}
\affiliation{Center for Gravitation and Cosmology, Yangzhou University, Yangzhou 225009, China}

\title{Beyond Coleman's Instantons}

\begin{abstract}
In the absence of gravity, Coleman's theorem states that the $O(4)$-symmetric instanton solution, which is regular at the origin and exponentially decays at infinity, gives the lowest action. Perturbatively, this implies that any small deformation from $O(4)$-symmetry gives a larger action.
In this letter we investigate the possibility of extending this theorem to the situation where the $O(4)$-symmetric instanton is singular, provided that the action is finite.  
In particular, we show a general form of the potential around the origin, which realizes a singular instanton with finite action.
We then discuss a concrete example in which this situation is realized, and analyze non-trivial anisotropic deformations around the solution perturbatively.
Intriguingly, in contrast to the case of Coleman's instantons, we find that there exists a deformed solution that has the same action as the one for the $O(4)$-symmetric solution up to the second order in perturbation.
Our result implies that there exist non-$O(4)$-symmetric solutions with finite action beyond Coleman's instantons, and gives rise to the possibility of the existence of a non-$O(4)$-symmetric instanton with a lower action.
\end{abstract}

\maketitle

\textit{Introduction}. The study of instanton solutions in quantum field theory has long been a cornerstone in understanding non-perturbative phenomena. 
Instantons, as classical solutions to the equations of motion in Euclidean space, offer physical insights into the tunneling process from a metastable (false) vacuum to a stable (true) vacuum~\cite{Coleman:1977py,Callan:1977pt}.
In a model with a single field $\Phi$, one employs the semiclassical approximation to express the decay rate $\Gamma$ per unit time per unit volume ($\mathcal{V}$) as
\begin{align}\label{eq:decay_rate}
    \Gamma/\mathcal{V}  = A\,e^{-B[\Phi]} \;,
\end{align}
where $A$ is the prefactor that contains $\hbar$ corrections,
$B[\Phi]$ is the bounce action defined in terms of the Euclidean action $S[\Phi]$ as
\begin{align}\label{eq:bounce_action}
    B[\Phi] \equiv S[\Phi] - S[\Phi_{\rm FV}] \;,
\end{align}
with $\Phi_{\rm FV}$ being the location of the false vacuum.  
Here we ignore quantum corrections and focus on the leading WKB approximation, i.e., semi-classical approximation.
Note that in the semiclassical limit one evaluates the action $S[\Phi]$ on a classical solution subjected to appropriate boundary conditions.
Indeed, it was mathematically proven in \cite{Coleman:1977th} that in the absence of gravity the $O(4)$-symmetric solutions result in the minimal bounce action 
under certain conditions. 
Those solutions were assumed to satisfy the boundary conditions,
\begin{align}\label{eq:con_coleman}
\Phi(\rho)\big|_{\rho \rightarrow \infty} = \Phi_{\rm FV} \;,  \quad \frac{\mathrm{d}\Phi(\rho)}{\mathrm{d}\rho} \bigg|_{\rho \rightarrow 0}= 0 \;,
\end{align}
where $\rho$ denotes the radial coordinate in 4-dimensional Euclidean space. 
The second condition in (\ref{eq:con_coleman}) guarantees the regularity of the solution at the center of true-vacuum bubble. 
We call Euclidean solutions that satisfy the condition (\ref{eq:con_coleman}) Coleman's instantons.
There have been many studies on the properties on various types of Coleman's instantons, e.g.~\cite{Fubini:1976jm,Linde:1981zj,Lee:1985uv,Tye:2009rb}.    
Recently, a class of (unbounded) potentials in an arbitrary spacetime dimension was identified, for which the Coleman instantons do not exist \cite{Mukhanov:2021rpp,Mukhanov:2021kat,Espinosa:2019hbm}.
See \cite{Mukhanov:2020pau,Mukhanov:2020wim,Mukhanov:2021ggy,Espinosa:2021qeo} for new instantons\footnote{The authors introduced the UV cutoff to make the action finite.} and pseudo bounces which do not satisfy the boundary conditions (\ref{eq:con_coleman}).
In this letter we investigate another case where $O(4)$-symmetric Euclidean solutions are singular but their action remains finite.\footnote{This situation is reminiscent of the Hawking-Turok instanton \cite{Hawking:1998bn} where gravity is taken into account. This instanton solution diverges as $\rho \rightarrow 0$ with finite action. 
See also \cite{Dine:2004uw} for a study of singular bounces in the context of bubble of nothing instantons.}
Namely, we consider $O(4)$-symmetric instantons that do not obey the second condition of (\ref{eq:con_coleman}), hence they are singular at the true-vacuum bubble. 
Also, similar to \cite{Mukhanov:2021rpp,Mukhanov:2021kat}, we are interested in an unbounded potential, which violates one of the assumptions for Coleman's theorem.
Moreover, we explicitly show that even if small and regular deformations are taken into account, the total action is still given by that computed on the $O(4)$-symmetric ones.
Here we do not include gravity.

Although gravitational effects are not included in the present analysis, our results already carry important implications for cosmology, for instance, for phase transitions in the early universe in the limit where the vacuum energy is much smaller than the Planck scale. 
In this regard, we view our work as a first step toward understanding one of the most fundamental processes in the Universe--phase transitions--which often play a crucial role throughout the cosmological history. 
To make our conclusions more robust, we need to take into account gravity, which will require a much more sophisticated approach. 
Therefore, the analysis in this paper is an initial step toward a study where gravity plays an important role.
We leave this to future work. 

\textit{Setup}. Let us start with a real scalar-field model with the Euclidean action,
\begin{align}\label{eq:action_Eu}
S = \int \mathrm{d}^4x \sqrt{g} \bigg[\frac{1}{2}\,g^{\mu\nu} \partial_\mu \Phi(x) \partial_\nu \Phi(x) + V(\Phi) \bigg] \;,
\end{align}
where $V(\Phi)$ is the potential. 
We consider the 4d flat Euclidean metric, 
\begin{align}\label{eq:flat_metric}
\mathrm{d}s^2 = \mathrm{d}\tau^2 + \mathrm{d}\vec{x}^2 = \mathrm{d}\rho^2 + \rho^2(\mathrm{d}\theta^2 + \sin^2\theta\,\mathrm{d}\Omega_2^2) \;, 
\end{align}
where $\tau$ is a Euclidean time, $\rho \equiv \sqrt{\tau^2 + \vec{x}^2}$ and $\mathrm{d}\Omega_2^2 \equiv \mathrm{d}\chi^2 + \sin^2 \chi\,\mathrm{d}\phi^2$.
Note that the model may admit solutions of $\Phi$ which are not $O(4)$-symmetric.
It may also admit solutions which are singular at $\rho=0$.

The equation of motion (EoM) of $\Phi$ reads 
\begin{align}
\Box \Phi - \frac{\mathrm{d}V(\Phi)}{\mathrm{d}\Phi} = 0 \;,
\label{eq:EOM_phi}
\end{align}
where $\Box \equiv g^{\mu\nu}\partial_\mu \partial_\nu$. 
Eq.~(\ref{eq:EOM_phi}) describes the dynamics of instanton solution with prescribed boundary conditions: one at the origin of true-vacuum bubble ($\rho = 0$) and the other at false vacuum ($\rho \rightarrow \infty$). 

In order to construct the potential $V(\Phi)$, we first consider the $O(4)$-symmetric instanton: $\Phi = \bar{\Phi}(\rho)$. 
Eq.~(\ref{eq:EOM_phi}) then reduces to 
\begin{align}
\bar{\Phi}'' + \frac{3}{\rho} \bar{\Phi}' - \frac{\mathrm{d}V}{\mathrm{d}\Phi}\bigg|_{\Phi = \bar{\Phi}} = 0\;, \label{eq:EOM_phi_zeroth}
\end{align}
where $' \equiv {\rm d}/{\rm d}\rho$. The corresponding action reads
\begin{align}\label{eq:action_zero}
S_0 =  2\pi^2 \int_0^\infty \mathrm{d}\rho~\rho^3~\bigg[\frac{1}{2}\bar{\Phi}'^2 + V(\bar{\Phi}) \bigg] \;,
\end{align}
where the factor $2\pi^2$ comes from the integrals over the angular variables.

Here we allow an instanton to be singular at the origin, or equivalently the true vacuum is located infinitely far away in field space, provided that the action remains finite.\footnote{Note that there could be a cuspy solution at the origin with the first derivative of the scalar field remains finite.}
From (\ref{eq:action_zero}), we see that if $\bar{\Phi}'^2(\rho)$ blows up faster than $1/\rho^3$ when $\rho \rightarrow 0$, the action will diverge. 
Therefore, the finiteness of the action yields the conditions on the behaviors of $\bar{\Phi}'$ and $\bar{\Phi}$ at the origin.

From the above reasoning, we find that the general form of the potential as $\rho \rightarrow 0$ with finite action is 
\begin{align}\label{eq:potential_gen_con}
    V(\Phi) = -\Lambda^4\exp\bigg(\frac{2\Phi}{\Phi_\star}\bigg) \;,
\end{align}
where $\Lambda$ and $\Phi_\star$ are constants. 
It is evident that Eq.~(\ref{eq:potential_gen_con}) leads to logarithmic divergence of $\bar{\Phi}$ at the origin, but it gives finite action.
Indeed, Eq.~(\ref{eq:potential_gen_con}) violates one of the assumptions of the theorem \cite{Coleman:1977th}, so that it does not apply in our case.
Notice that the potential (\ref{eq:potential_gen_con}) has not been considered in the literature, and here we do not need to introduce a UV cutoff since our integral over $\rho$ is finite.
See \cite{Sasaki:2025nzy} for discussion of other types of singularities of $\bar{\Phi}$.
Note also that the potential (\ref{eq:potential_gen_con}) is generic in $d > 2$ dimensions.

\textit{Small anisotropic deformation}. Let us analyze a small anisotropic deformation $\delta \Phi$ of the $O(4)$-symmetric instanton,
\begin{align}\label{eq:ansatz}
    \Phi(x^\mu) = \bar{\Phi}(\rho) + \epsilon\, \delta\Phi(\rho, \vec{\theta}\,) \;,
\end{align}
where $\epsilon$ is the anisotropic parameter satisfying $\epsilon\ll1$.
Here we emphasize that we focus on deformations that satisfy the classical equations of motion.
Notice that $\delta\Phi$ can generally depend on both $\rho$ and the angular variables, $\vec{\theta} = \{\theta, \chi, \phi\}$.  
We decompose $\delta\Phi(\rho, \vec{\theta}\,)$ as
\begin{align}
    \delta\Phi(\rho, \vec{\theta}\,) = \sum_{L,M} A_{LM}(\rho)\,Y_L^M(\vec{\theta}\,) \;, \label{eq:sol_PHi1}
\end{align}
where the function $Y_L^M(\vec{\theta}\,)$ is the 3-dimensional spherical harmonics \cite{doi:10.1063/1.527513}, obeying
\begin{align}
    \Delta_{S^{3}} Y_L^M(\vec{\theta}\,) = -L(L + 2) Y^M_L(\vec{\theta}\,) \;,
\end{align}
with $\Delta_{S^{3}}$ the Laplacian of a 3-sphere and $M$ a multi-index characterizing the magnetic quantum numbers $\{m_\phi, m_\chi, m_\theta\}$ satisfying $|m_\phi| \leq m_\chi \leq L \equiv m_\theta$.

The radial component $A_L(\rho)$ satisfies
\begin{align}\label{eq:A(rho)}
    A''_L + \frac{3}{\rho} A'_L - \bigg[ \frac{\de ^2 V}{\de\Phi^2}\bigg|_{\bar{\Phi}} + \frac{L(L + 2)}{\rho^2} \bigg]A_L = 0 \;,
\end{align}
where we omit the index $M$. 
Eq.~(\ref{eq:A(rho)}) determines the dynamics of fluctuations $A_L$ on the background $\bar{\Phi}$.
Notice that the modes with $L = 0$ and $L = 1$ are not physical, see \cite{Sasaki:2025nzy} for detailed discussion.\footnote{Note that if the theory has conformal symmetry, there could exist regular $L=0$ deformations. However, a non-vanishing second derivative of the potential at the false vacuum violates this symmetry. Hence, there is no regular $L=0$ deformations.}
Here and below we ignore $M$ dependence of the perturbations since the equations do not depend on $M$.
Inserting (\ref{eq:ansatz}) into (\ref{eq:action_Eu}), one arrives at the action up to $\mathcal{O}(\epsilon^2)$,
\begin{align}\label{eq:action_Phi1}
S = S_0[\bar{\Phi}] + \epsilon^2 S_2[\delta\Phi] \;,     
\end{align}
where $S_0$ is given by Eq.~(\ref{eq:action_zero}) and 
\begin{align}\label{eq:action_Phi2}
S_2 =\frac{1}{2} \int \mathrm{d}^4x \sqrt{g}\,\bigg[(\partial \delta\Phi)^2 + \frac{\de^2V}{\de\Phi^2}\bigg|_{\bar{\Phi}} \delta \Phi^2 \bigg]   \;.    
\end{align}
Note that we have performed the integration by parts and used Eq.~(\ref{eq:EOM_phi_zeroth}) to obtain (\ref{eq:action_Phi1}).
(The action $S_2$ is the leading-order correction to the $O(4)$-symmetric on-shell action $S_0$.)
Note also that the effective mass for $A_L$, i.e., $\de ^2V/\de\Phi^2$ evaluated on $\bar{\Phi}$ is only a function of $\rho$; the only non-trivial angular dependence appears through $\delta\Phi(\rho, \vec{\theta}\,)$. 
Hence, using the solution (\ref{eq:sol_PHi1}) in Eq.~(\ref{eq:action_Phi2}) gives
\begin{align}
    S_2 = \sum_{L \geq 2} \int \mathrm{d}\rho\, \mathcal{L}_{2L} \label{eq:second_actio_AL} \;,
\end{align}
where
\begin{align}\label{eq:integrand_A_L}
    \mathcal{L}_{2L} \equiv \frac{\rho^3}{2} \bigg\{A_L'^2 + \bigg[\frac{L(L+2)}{\rho^2}
    + \frac{\de^2 V}{\de \Phi^2}\bigg|_{\bar{\Phi}}\bigg] A_L^2\bigg\} \;,
\end{align}
and we have used the orthogonality condition for spherical harmonics.
For a regular $A_L$, Eq.~(\ref{eq:second_actio_AL}) can be integrated as
\begin{align}
   S_2 = \pi^2 \sum_{L \geq 2} \rho^3 A_L(\rho) A'_L(\rho) \bigg|_{\rho = 0}^{\rho = \infty} \;, \label{eq:s_2_phi_1}
\end{align}
where we have performed an integration by parts and used Eq.~(\ref{eq:A(rho)}).
Hence, for regular $A_L$, the action $S_2$ is solely determined by the behaviors of $A_L$ at the origin and infinity, which give a finite value of $S_2$.
In this sense, our model is beyond Coleman theorem \cite{Coleman:1977th} since the $O(4)$-symmetric solution $\bar{\Phi}$ is assumed to be singular at $\rho = 0$ and the potential is not admissible everywhere.
It may be interesting to have a rigorous proof of the least action with a singular instanton, we leave this to future work.
In the following, we present a concrete example which explicitly realizes the case we are interested in. 

\begin{figure}[t]
\includegraphics[width=0.45\textwidth]{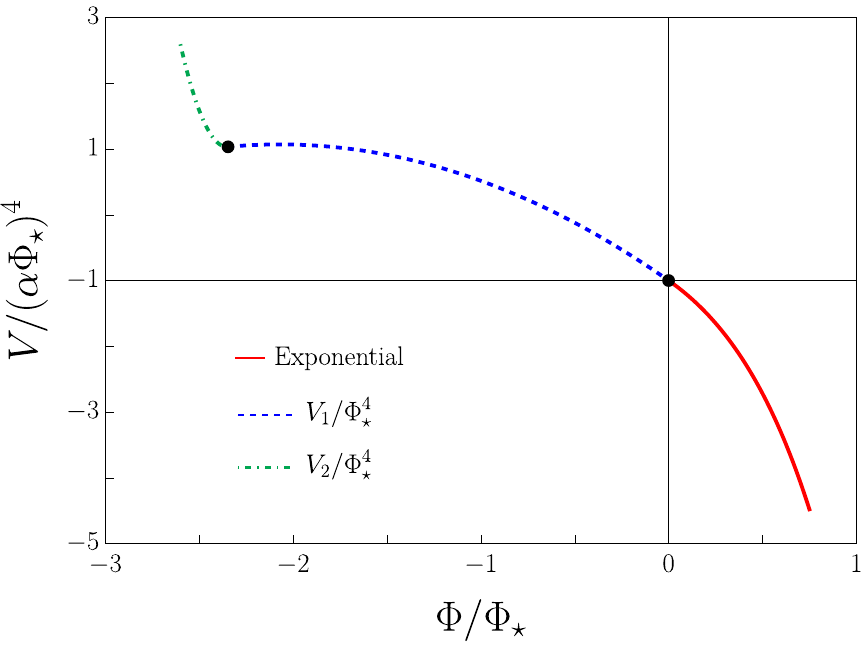}
\caption{~$V(\Phi)/(\alpha\Phi_\star)^4$ in (\ref{eq:piecewise_potential}) as a function of $\Phi$ with $m_1 = 0.98\Phi_\star$, $m_2 = 7.11 \Phi_\star$ and $\Phi_{\rm M} = -2.35\Phi_\star$. 
These values of parameters are chosen such that the matching conditions for Eqs.~(\ref{eq:bg_exp})--(\ref{eq:sol_phi_0_V2}) are satisfied.
The solid red line represents the exponential potential, while the blue dashed and the green dot-dashed lines refer to the potentials $V_1$ and $V_2$ in Eq.~(\ref{eq:piecewise_potential}) respectively. The two black points in the plot refer to the two matching locations at $\Phi = 0$ and at $\Phi = \Phi_2 = -2.34\Phi_\star$. The false vacuum for this particular choice of parameters is located at $\Phi_{\rm M} = -2.35\Phi_\star$.}
\label{fig:piecewise_potential} 
\end{figure}

\textit{Example: Piecewise quadratic potential}. We study an example with the potential
\begin{align}\label{eq:piecewise_potential}
    \frac{V(\Phi)}{\alpha^4} = \left\{
\begin{matrix}
-\Phi_{\star}^4 \exp(2 \Phi/\Phi_\star)\;, & \Phi \geq 0 \\
V_{1}(\Phi) \;, & \Phi_2 \leq \Phi \leq 0 \\
V_{2}(\Phi) \;, & \Phi \leq \Phi_2 
\end{matrix}
\right. \;,
\end{align}
where 
\begin{align}
    V_{1}(\Phi) &\equiv -\frac{1}{2}m_1^2(\Phi - \Phi_{\rm P})^2 - \Lambda_1^4 \;, \\
    V_{2}(\Phi) &\equiv \frac{1}{2}m_2^2(\Phi - \Phi_{\rm M})^2 - \Lambda_2^4 \;, 
\end{align}
where $\Lambda$, $m_1$, $m_2$, $\Phi_{\rm P}$, $\Phi_{\rm M}$, $\Lambda_1$, $\Lambda_2$, $\Phi_\star$ are parameters with mass dimension,
$\alpha$ is an overall dimensionless parameter.
Notice that $\Phi_{\rm P}$ and $\Phi_{\rm M}$ denote the locations of the maximum and the minimum (false vacuum) in $V_1$ and $V_2$, respectively.
Imposing the continuity conditions for the potential and its first derivative at $\Phi = 0$ and $\Phi = \Phi_2$, one can fix the parameters $\{\Lambda_1, \Lambda_2, \Phi_{\rm P}, \Phi_2\}$ in terms of $\{m_1, m_2 ,\Phi_{\rm M}\}$ such that
\begin{equation}\label{paracon}
\begin{aligned}
     \Phi_{\rm P} &= -\frac{2\Phi_\star^3}{m_1^2} \;, \\
    \Phi_2 &= -\frac{2 \Phi_\star^3}{m_1^2 + m_2^2}\bigg(1 - \frac{m_2^2\Phi_{\rm M}}{2\Phi_\star^3}\bigg) \;, \\
    \Lambda_1^4 &=  \Phi_\star^4 \bigg(1 - \frac{2\Phi_\star^2}{m_1^2}\bigg) \;, \\ 
    \Lambda_2^4 &= -\frac{2\Phi_\star^6 }{m_1^2+m_2^2} \bigg[1 - \bigg(1+\frac{m_2^2\Phi_{\rm M}^2}{2\Phi_\star^4}\bigg)\frac{m_1^2}{2 \Phi_\star^2} \\
    & \hspace{4.2mm} - \bigg(1 + \frac{2\Phi_{\rm M}}{\Phi_\star}\bigg)\frac{m_2^2}{2\Phi_\star^2}\bigg]\;.
\end{aligned}
\end{equation}
Moreover, the location of the false vacuum must fall into the regime of $V_2$, which implies that $\Phi_{\rm M} < \Phi_2 < 0$. This gives the condition\footnote{One can straightforwardly verify that the condition (\ref{eq:con_m1}) automatically leads to $\Phi_2 \leq \Phi_{\rm P} < 0$.}, 
\begin{align}\label{eq:con_m1}
    2 + \frac{m_1^2 \Phi_{\rm M}}{ \Phi_\star^3} < 0 \;.
\end{align}
For illustrative purposes, we plot the potential (\ref{eq:piecewise_potential}) in Fig.~\ref{fig:piecewise_potential}.

\begin{figure}[t]
\includegraphics[width=0.45\textwidth]{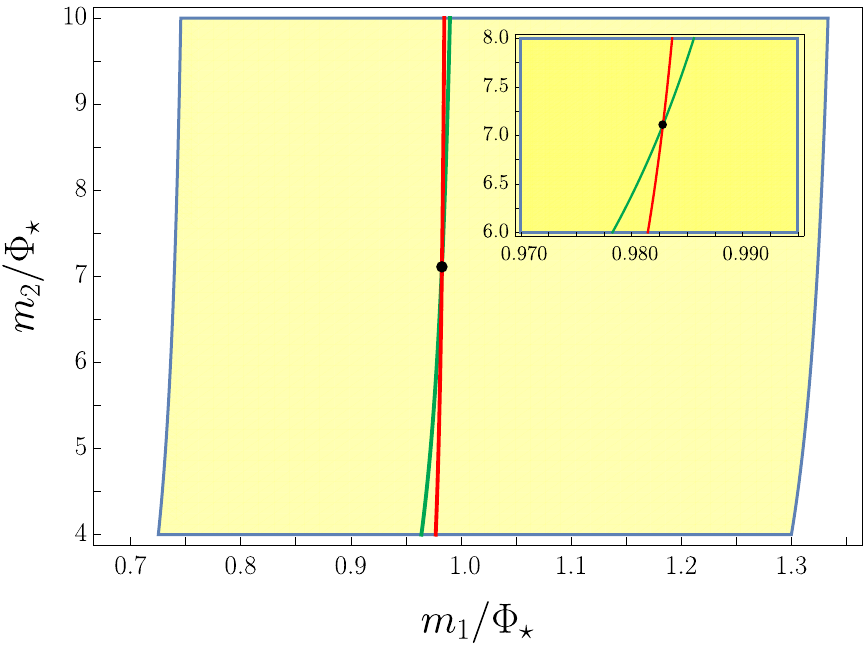}
\caption{~Parameter space of $m_1/\Phi_\star$ and $m_2/\Phi_\star$.
The condition (\ref{eq:con_bg}) is satisfied in the yellow region.
The green line refers to the condition $\bar{\Phi}(\tilde{\rho}_2) = \Phi_2$, the red curve corresponds to the condition $c_8 = 0$.
We choose $\tilde{\rho}_2 = 5$ and $\alpha = 0.5$.
The black dot is $\{0.98,7.11\}$, for which the fluctuation is regular everywhere (their corresponding $\Phi_2$ and $\Phi_{\rm M}$ are respectively given by $\Phi_2 = -2.34\Phi_\star, \Phi_{\rm M} = -2.35\Phi_\star$). The small figure explicitly shows the intersection point.}
\label{fig:contour_plot_piecewise} 
\end{figure}

At the background level, in each regime of the potential (\ref{eq:piecewise_potential}) $\bar{\Phi}(\tilde{\rho})$ can be obtained analytically without the thin-wall approximation. 
In the exponential-potential regime, solving Eq.~(\ref{eq:EOM_phi_zeroth}) gives 
\begin{align}\label{eq:bg_exp}
    \bar{\Phi}(\tilde{\rho}) = -\Phi_\star\log(\tilde{\rho}) \,;~0<\tilde{\rho}\leqslant1 \,,
\end{align}
where $\tilde{\rho} \equiv \alpha^2 \Phi_\star \rho$. 
Notice that in (\ref{eq:bg_exp}) we focus on the singular behavior, so that the regular behavior is disregarded.\footnote{Of course, once the condition $\Phi'_0 \rightarrow \infty$ at $\rho = 0$ is imposed, $\bar{\Phi}$ is uniquely determined by Eq.~(\ref{eq:bg_exp}).}
In the $V_1$ regime the solution reads 
\begin{align}
    \bar{\Phi} (\tilde{\rho}) = & - \frac{2\Phi_\star^3}{m_1^2} + \frac{c_1}{\tilde{\rho}} J_1\big(\frac{m_1}{\Phi_\star}\tilde{\rho}\big) + \frac{c_2}{\tilde{\rho}}Y_1\big(\frac{m_1}{\Phi_\star} \tilde{\rho}\big) \,; \nonumber \\
    & 1\leqslant\tilde{\rho}\leqslant\tilde{\rho}_2 \,,
\end{align}
where $J_1$ and $Y_1$ are Bessel functions of the first and second kinds respectively, and $c_1$, $c_2$ are constants.
In the $V_2$ regime we have 
\begin{align}\label{eq:sol_phi_0_V2}
    \bar{\Phi}(\tilde{\rho}) = \Phi_{\rm M} + \frac{c_3}{\tilde{\rho}} K_1\big(\frac{m_2}{\Phi_\star}\tilde{\rho}\big) + \frac{c_4}{\tilde{\rho}}I_1\big( \frac{m_2}{\Phi_\star}\tilde{\rho}\big) \,;~\tilde{\rho}\geqslant\tilde{\rho}_2 \,,
\end{align}
where $K_1$ and $I_1$ are modified Bessel functions of the first and second kinds respectively\footnote{The Bessel functions $K_1(z)$, $I_1(z)$ behaves as $e^{-z}$ and $e^{z}$ as $z \rightarrow \infty$.}, and $c_3$, $c_4$ are constants.
In order for $\bar{\Phi}$ asymptotically goes to $\Phi_{\rm M}$ as $\tilde{\rho} \rightarrow \infty$, we impose $c_4 = 0$.
Then, using the continuity conditions at $\tilde{\rho} = 1$ ($\bar{\Phi} = 0$) the constants $c_1$ and $c_2$ are fixed in terms of $m_1$. 
At the second matching point $\tilde{\rho} = \tilde{\rho}_2$ ($\bar{\Phi} = \Phi_2$), using the continuities of both $\bar{\Phi}$ and $\bar{\Phi}'$ yields
\begin{align}
    \bigg(2 + \frac{m_1^2 \Phi_{\rm M}}{\Phi_\star^3}\bigg) H_1 + H_2 = 0 \;,
\end{align}
where $H_1$ and $H_2$ are functions of $m_1$, $m_2$, and $\tilde{\rho}_2$. Hence, $\Phi_{\rm M}$ is also fixed by these parameters.
Therefore, inserting the expression for $\Phi_{\rm M}$ into (\ref{eq:con_m1}), we obtain the inequality, 
\begin{align}\label{eq:con_bg}
    H_3\left(\frac{m_1}{\Phi_\star}, \frac{m_2}{\Phi_\star}, \tilde{\rho}_2 \right) \leq 0 \;.
\end{align}
Here we omit the full expressions of $H_1$, $H_2$ and $H_3$ due to their length.
In addition to (\ref{eq:con_bg}), we identify the solution \eqref{eq:sol_phi_0_V2} at $\tilde{\rho} = \tilde{\rho}_2$ with $\Phi_2$ in (\ref{paracon}), i.e.,
\begin{align}\label{eq:con_bg_Phi_2}
\bar{\Phi}(\tilde{\rho}_2) = \Phi_2 \,.
\end{align}
For $\tilde{\rho}_2 > 1$, the conditions (\ref{eq:con_bg}) and (\ref{eq:con_bg_Phi_2}) must be satisfied so that the background solution can be smoothly connected and asymptotically goes to the false vacuum $\Phi_{\rm M}$. 
In Fig.~\ref{fig:contour_plot_piecewise} we demonstrate the existence of such solutions by the green line in the yellow region.
After fixing $\tilde{\rho}_2$ and imposing the above condition, the remaining parameters are $m_1$ and $\alpha$.

Here, we point out that our singular bounce solution is unique.
The reason is as follows. 
As for our model, since the potential is quadratic at $\Phi < 0$, the solution is known to be unique once the boundary condition is fixed. 
As for the exponential potential at $\Phi>0$, the equation is the ordinary 2nd order differential equation. 
Therefore, the value and its derivative at $\Phi=0$ uniquely determines the solution at $\Phi>0$.
This implies that our solution is unique, i.e, there is no regular bounce solutions.

Having determined the background solution $\bar{\Phi}$, now we turn to analyze the small deformation $\delta\Phi(\rho, \vec{\theta}\,)$ in \eqref{eq:ansatz}. In order to solve its radical components $A_L$, we introduce a dimensionless variable $f_L \equiv \tilde{\rho}^{3/2} A_L/\Phi_\star$.  
In the exponential regime of potential \eqref{eq:piecewise_potential}, solving (\ref{eq:A(rho)}) gives
\begin{align}\label{eq:sol_f_expo}
    f_L(\tilde{\rho}) = a_1 \tilde{\rho}^{\frac{1}{2} + \sqrt{j}} + a_2 \tilde{\rho}^{\frac{1}{2} -\sqrt{j}}\,;~0<\tilde{\rho}\leqslant1\,,
\end{align}
where $a_1$, $a_2$ are constants and $j \equiv (L - 1)(L + 3)$. 
Here we focus on the regular behavior at the origin, i.e., $a_2 = 0$, in order for the action to be finite.
Note that this choice excludes the possibility that the solution behaves as $1/\sqrt{\tilde{\rho}}$.\footnote{This is the only singular behavior with finite action, see Eq.~(\ref{eq:integrand_A_L}).}
In the $V_1$ regime we obtain
\begin{align}\label{eq:sol_first_f_L}
    f_L(\tilde{\rho}) =& \sqrt{\tilde{\rho}}\left[c_5 J_{L+1}\big(\frac{m_1}{\alpha^2\Phi_\star}\tilde{\rho}\big) +  c_6 Y_{L+1}\big(\frac{m_1}{\alpha^2\Phi_\star}\tilde{\rho}\big) \right]\,;\nonumber\\
    & 1\leqslant\tilde{\rho}\leqslant\tilde{\rho}_2\;,
\end{align}
where $c_5$ and $c_6$ are constants. 
Similarly, the solution in the $V_2$ regime reads 
\begin{align}\label{eq:sol_second_f_L}
    f_L(\tilde{\rho}) = &\sqrt{\tilde{\rho}}\left[c_7 K_{L+1}\big(\frac{m_2}{\alpha^2\Phi_\star}\tilde{\rho}\big) +  c_8 I_{L+1}\big(\frac{m_2}{\alpha^2\Phi_\star}\tilde{\rho}\big) \right] \,;\nonumber \\
    & \tilde{\rho}\geqslant\tilde{\rho}_2 \,,
\end{align}
where $c_7$ and $c_8$ are constants. 
Note that the $\alpha$ dependence in the solutions (\ref{eq:sol_first_f_L}) and (\ref{eq:sol_second_f_L}) can be understood from a constant shift in $\Phi$, $\Phi \to \Phi + const.$, which introduces a new multiplicative factor in front of the exponential potential. 
Fixing this shift parameter directly affects the term ${\rm d}^2V/{\rm d}\Phi^2|_{\bar{\Phi}}$ in the piecewise quadratic potential regimes, leading to $m_1 \to m_1/\alpha^2$ and $m_2 \to m_2/\alpha^2$.
Then, using the matching conditions for $f_L(\tilde{\rho})$ and its first derivative at $\tilde{\rho} = 1$ one can fix the constants $c_5$ and $c_6$ in terms of $a_1$, $m_1$ and $\alpha$.
Furthermore, imposing the matching conditions at $\tilde{\rho} = \tilde{\rho}_2$ one can determine $c_7$ and $c_8$ in terms of $\alpha$, $m_1$, $m_2$, $\tilde{\rho}_2$ and $a_1$. 
Finally, we require the solution (\ref{eq:sol_second_f_L}) to be regular as $\tilde{\rho} \rightarrow\infty$, this fixes $c_8(m_1, m_2, \tilde{\rho}_2, a_1, \alpha) = 0$. 
For given values of $\tilde{\rho}_2$, $\alpha$ and $a_1$, this condition  provides a relation between $m_1$ and $m_2$.
Note that the coefficient $a_1$ can be set to unity without loss of generality.
\begin{figure}[t]
\includegraphics[width=0.45\textwidth]{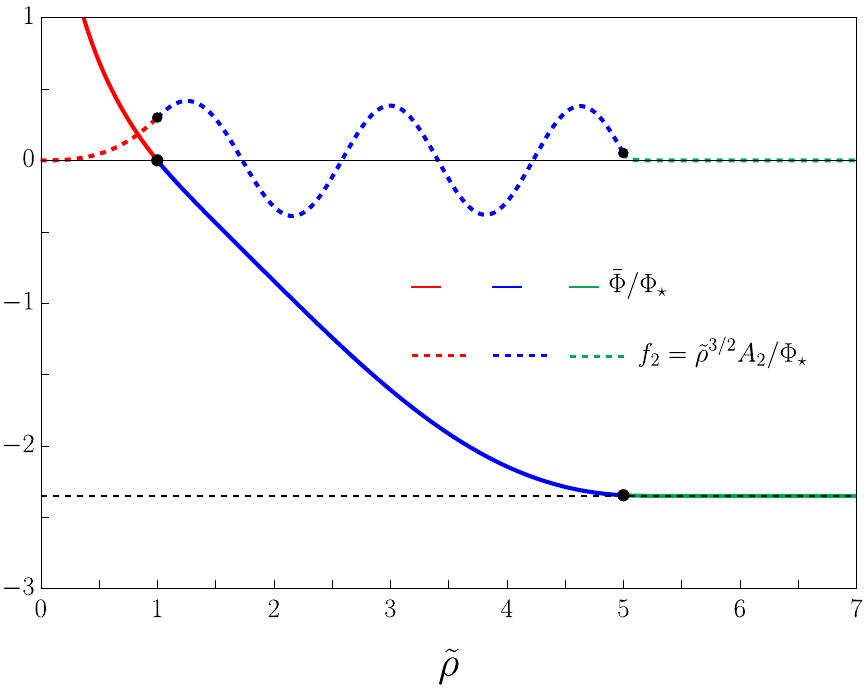}
\caption{~The solutions for $\bar{\Phi}$ (solid line) and $f_2(\tilde{\rho})$ (dashed line). 
The red, blue and green colors refer to the solutions in the exponential potential, $V_1$ and $V_2$ regimes, respectively.
We set $\alpha = 0.5$, $m_1 = 0.98\Phi_\star$ and $m_2 = 7.11\Phi_\star$, so that $\Phi_2 = -2.34\Phi_\star$ and $\Phi_{\rm M} = -2.35\Phi_\star$.
The two matching points (black dots) are located at $\tilde{\rho} = 1$ and $\tilde{\rho}_2 = 5$.}
\label{fig:fluctuation_piecewise} 
\end{figure}

In Fig.~\ref{fig:contour_plot_piecewise} we illustrate the existence of both $\bar{\Phi}$ and regular deformations by the intersection point where the conditions [$c_8 = 0$, Eqs.~(\ref{eq:con_bg}) and (\ref{eq:con_bg_Phi_2})] are satisfied.
To be specific, the degrees of freedom (dofs) of the parameters of our model are 5: ($m_1,m_2,\Phi_{\rm M},\Phi_\star,\alpha$). 
The conditions $\bar{\Phi}(\rho_2) = \Phi_2$ and $c_4 = 0$ (exponentially decaying of $\bar{\Phi}$ as $\rho \to \infty$) reduce $5 - 2 = 3$ dofs. 
All the coefficients $\{c_5,c_6,c_7,c_8\}$ of the perturbation solution are completely determined when imposing the matching conditions both at $\tilde{\rho} = 1$ and $\tilde{\rho} = \tilde{\rho}_2$. 
Therefore, requiring $c_8 = 0$, we are left with $3-1 = 2$ dofs. 
We take them to be $\tilde{\rho}_2$ and $\alpha$.
For other values of $\tilde{\rho}_2$ and $\alpha$ one needs to check the existence of the solutions case by case. 

For completeness, in Fig.~\ref{fig:fluctuation_piecewise} we plot both the solutions $\bar{\Phi}$ (solid line) and $f_2$ (dashed line) as a function of $\tilde{\rho}$. 
Note that in the plot we appropriately rescale $f_2$ with the parameter $a_1$.
We see that $\bar{\Phi}$ is singular as $\tilde{\rho} \to 0$, while it goes to $\Phi_{\rm M}$ as $\tilde{\rho} \to \infty$.
Moreover, the solution $f_2(\tilde{\rho})$ is regular everywhere, especially at the origin and infinity.
Therefore, in this model there exists the regular solution for small deformations around the singular instanton. 

More importantly, the deformation is regular everywhere. Inserting the solution given by \eqref{eq:sol_f_expo} and \eqref{eq:sol_second_f_L} with $a_2=c_8=0$ into (\ref{eq:s_2_phi_1}), we find $S_2=0$.
Therefore, the total action (\ref{eq:action_Phi1}) is the same as the one evaluated for the $O(4)$-symmetric instanton. 
In other words, a small and regular deformation can be regarded as a zero mode on the $O(4)$-symmetric solutions.
Our result implies that there exist non-$O(4)$-symmetric solutions with finite action beyond Coleman's instantons.

\textit{Discussion}. We studied a possibility of extending the theorem \cite{Coleman:1977th} to the case where the $O(4)$-symmetric instanton solution exhibits a singular behavior at $\rho = 0$. 
In particular, we considered the case in which the solution does not satisfy Coleman's condition (\ref{eq:con_coleman}).
We further investigated a small anisotropic deformation around the singular $O(4)$-symmetric instanton. 
We proposed a concrete example with a piecewise potential given by (\ref{eq:piecewise_potential}).
In our separate work \cite{Sasaki:2025nzy} we discuss the generic form of the instanton potential, and other types of singular behaviors in addition to logarithmic divergence.
We then analyzed the solutions for $\bar{\Phi}$, that is singular at $\rho = 0$ and goes to false vacuum as $\rho \rightarrow \infty$.
Finally, we studied the dynamics of small anisotropic deformation and found that there exists a small and regular deformation around the singular $O(4)$-symmetric solution, which does not contribute to the total bounce action. 
Our result implies that there exist non-$O(4)$-symmetric solutions with finite action beyond Coleman's instantons, and gives rise to the possibility of the existence of a non-$O(4)$-symmetric instanton with a lower action.
To clarify this point, it is necessary to make a non-linear analysis, which is a future issue.
In any case, our result opens several possibilities to study other cases, where the Coleman's theorem may not apply.

First of all, it could be that the quantum corrections to a singular instanton with finite action diverge. This issue also certainly deserves further study.
Second, it would be interesting to explore whether the so-called pseudo bounces \cite{Espinosa:2019hbm,Espinosa:2021qeo} give lower or higher action than the one in our case, and study their implications.\footnote{It could be that some off-shell configurations lead to lower actions, but they may correspond to subcritical bubbles.} 
Moreover, as discussed in a paragraph above Eq.~(4), it is necessary to extend our analysis to the case with gravity.
The results presented here may therefore be regarded as a first step toward a sophisticated study, in which gravity plays an important role.
In connection with this, see e.g. \cite{Cohn:1998et,Gratton:1999ya,Dunne:2006bt,Ai:2023yce,Oshita:2023pwr,Oshita:2021aux}.

\textit{Acknowledgements}. We thank S.~Mukohyama, N.~Oshita, R.~Saito and K.~Takahashi for useful discussions. We thank J.~R.~Espinosa for pointing out an error in the first version of the paper. 
This work is supported in part by JSPS KAKENHI Nos. JP20H05853 and JP24K00624.
M.S. and V.Y. are supported by World Premier International Research Center Initiative (WPI Initiative), MEXT, Japan. 
V.Y. is supported in part by grants for development of new faculty staff, Ratchadaphiseksomphot Fund, Chulalongkorn University and by the NSRF via the Program Management Unit for Human Resources \& Institutional Development, Research and Innovation Grant No.\ B39G680009.
Y.Z. is supported by  the Fundamental Research Funds for the Central Universities, and by the Project 12475060 and 12047503 supported by NSFC, Project 24ZR1472400 sponsored by Natural Science Foundation of Shanghai.

	{}
\bibliographystyle{utphys}
\bibliography{bib_v4.bib}

\end{document}